\documentclass[10pt,letterpaper,twocolumn]{article} 

\usepackage{ol2}
\usepackage[draft]{hyperref}
\usepackage{amsmath}

\begin{document}

\twocolumn[ 

\title{A photonic crystal nanocavity laser in an optically very thick slab}


\author{Se-Heon Kim,$^{1,2,*}$ Jingqing Huang,$^{1,2}$ and Axel Scherer$^{1,2}$}

\address{
$^1$Department of Electrical Engineering, California Institute of Technology, Pasadena, CA 91125, USA
\\
$^2$Kavli Nanoscience Institute, California Institute of Technology, Pasadena, CA 91125, USA
\\
$^*$Corresponding author: seheon@caltech.edu
}

\begin{abstract}
A photonic crystal (PhC) nanocavity formed in an optically {\em very thick} slab can support reasonably high-$Q$ modes for lasing. Experimentally, we demonstrate room-temperature pulsed lasing operation from the PhC dipole mode emitting at 1324 nm, which is fabricated in an InGaAsP slab with thickness ($T$) of 606 nm. Numerical simulation reveals that, when $T \geq 800$ nm, over 90 \% of the laser output power couples to the PhC slab modes, suggesting a new route towards an efficient in-plane laser for photonic integrated circuits. 
\end{abstract}


 ] 

An optically {\em thin} dielectric slab with photonic crystal (PhC) air-holes has been a versatile platform for designing various high-$Q$ cavities.\cite{Painter99} Thickness ($T$) of the PhC slab is often chosen to maximize the size of the photonic band gap (PBG),\cite{S_G_Johnson_99} which is approximately equal to half effective wavelength of the cavity resonance. For designing a PhC slab laser emitting at 1.3 $\mu{\rm m}$, this thickness consideration requires that $T$ should be about 250 nm. 

In this letter, we show that even a {\em very thick} slab can support sufficiently high-$Q$ cavity modes for lasing. Once we are free from the thickness constraint, design of a current-injection type laser becomes more feasible; we can employ a vertically varying $p$-$i$-$n$ structure along with a current confinement aperture as has been done for vertical-cavity surface-emitting lasers.\cite{Iga08} Furthermore, as will be shown below, we can build an efficient {\em in-plane} emitting laser, where most of the laser emission couples to the two-dimensional (2D) Bloch modes\cite{S_G_Johnson_99} in the PhC slab.

\begin{figure}[b]
\centering\includegraphics[width=6.5cm]{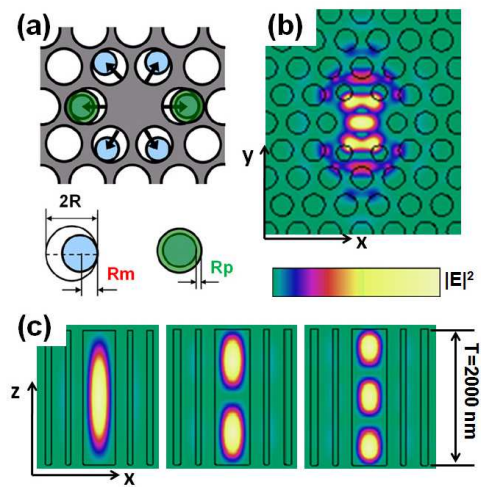}
\caption{\label{fig:fig1} (a) Design of the modified dipole cavity. (b,c) FDTD simulations for the dipole mode in a PhC slab with $T$ = 2,000 nm. (b) Top-down view of the electric-field intensity ($|{\bf E}|^2$) profile and (c) cross-sectional views of $|{\bf E}|^2$ of the fundamental, first-order, and second-order slab modes.}
\end{figure}

We begin with numerical simulations using the finite-difference time-domain (FDTD) method. We adopt the widely-used modified single-cell cavity design,\cite{Kim_PRB_06} and investigate the PhC dipole mode as shown in Fig.~\ref{fig:fig1}. We assume $T$ and the lattice constant ($a$) are 2,000 nm and 305 nm, respectively. The refractive index of the slab is assumed to be 3.4. Other structural parameters are as follows:\cite{Kim_PRB_06} the background hole radius ($R$) = $0.35 a$, the modified hole radius ($Rm$) = $0.25 a$, and the hole radius perturbation ($Rp$) = $0.05 a$. It should be noted that the in-plane PBG\cite{S_G_Johnson_99} is completely closed at $T \approx 1.5 a$ for a PhC slab with $R = 0.35 a$. However, it is interesting that we can still find several resonant modes that seem to be well confined within the defect region as shown in Fig.~\ref{fig:fig1}(b),(c). In fact, these modes have the same {\em transverse} mode profile while the number of intensity lobes along the $z$ direction varies from 1 to 3. Therefore, these modes originate from the slab resonance between the top and bottom surfaces, which can act as reflectors due to the relatively high refractive index of the slab.  We summarize various optical characteristics of the dipole modes in a slab with $T$ = 2,000 nm including $Q$, emission wavelength $\lambda$, and mode volume $V$, in Table~\ref{table:table1}.\cite{Kim_PRB_06} In particular, $Q_{\rm tot}$\cite{Qfactor} of the fundamental mode is over 5,000. It should be noted that a similar thick slab design was proposed by Tandaechanurat {\em et al.} with a special focus on a PhC cavity in a $T = 1.4 a$ slab.\cite{Tandaechanurat08}    

\begin{figure}[t]
\centering\includegraphics[width=7.5cm]{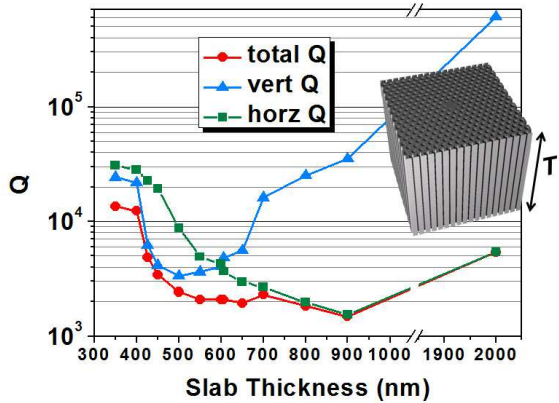}
\caption{\label{fig:fig2} $Q$ of the fundamental dipole mode as a function of slab thickness, where we fix the $x$-$y$ simulation domain size to be $16 a \times 16 a$}
\end{figure}

To gain further insight into the loss mechanism, in Fig.~\ref{fig:fig2}, we calculate $Q_{\rm tot}$, $Q_{\rm vert}$, and $Q_{\rm horz}$\cite{Painter99, Qfactor} as a function of $T$, where $a$ is varied to keep the emission wavelength at 1.3 $\mu{\rm m}$. First, let us focus on $Q_{\rm vert}$. In the case of a {\em thin} slab PhC cavity, $Q_{\rm vert}$ depends strongly on $Rm$ and $Rp$\cite{H_Y_Ryu_03} and $Q_{\rm vert}$ of the dipole mode can be as high as $\sim$ 15,000.\cite{Kim_PRB_06} Indeed, when $T \leq$ 400 nm, $Q_{\rm vert}$ is in the range of 10,000. However, when $T \geq$ 500 nm, $Q_{\rm vert}$ increases almost exponentially as $T$ increases. We obtain a surprisingly high $Q_{\rm vert}$ of $6 \times 10^5$ at $T$ = 2,000 nm, implying the existence of a certain highly-efficient vertical confinement mechanism, which will be clarified later. On the other hand, the in-plane confinement mechanism is not very effective as expected, because the PBG is closed for $T > \sim$ 450 nm. However, $Q_{\rm horz}$ can be brought up to $\sim$ 5,500 at $T$ = 2,000 nm and $Q_{\rm tot}$ is usually limited by $Q_{\rm horz}$ at large $T$. This large difference between $Q_{\rm vert}$ and $Q_{\rm horz}$ implies that most of the photons generated inside the cavity will leak into the PhC slab; at $T$ = 2,000 nm, over 99 \% (horizontal emission efficiency, $\eta_{\rm horz} = 1 - Q_{\rm tot}/Q_{\rm vert}$) of the total number of photons will be funneled through the PhC slab. $\eta_{\rm horz}$ is over 90 \% when $T \geq$ 800 nm. This behavior is completely opposite to the case of a {\em thin} slab cavity, where $Q_{\rm horz}$ can increase indefinitely by simply adding more layers of PhC barriers, therefore $Q_{\rm tot}$ is limited by $Q_{\rm vert}$. 

\begin{table}
  \centering
  \caption{\bf \label{table:table1}Optical properties of the higher-order slab modes}
   \begin{tabular}{ccccc} \\ \hline
    {} & $\lambda$(nm) & $Q_{\rm tot}$ & $Q_{\rm vert}$ & $V (\lambda/n)^3$ \\ \hline
    Fundamental & 1,324 & 5,392 & 6.1$\times 10^5$ &  2.45 \\
    1st-order & 1,305 & 1,582 & 41,600 &  2.65 \\
    2nd-order & 1,275 & 755 & 27,900 &  2.86 \\ \hline
  \end{tabular}
\end{table}

\begin{figure}[t]
\centering\includegraphics[width=8.4cm]{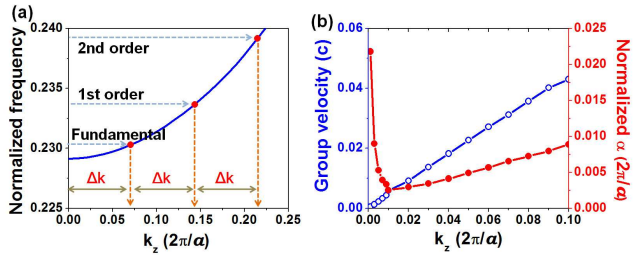}
\caption{\label{fig:fig3} (a) Waveguide dispersion along the $z$ direction for the dipole mode. The normalized frequencies of the three dipole resonant modes shown in Fig.~\ref{fig:fig1}(c) are overlaid on the dispersion curve. (b) Group velocity ($V_g$) and waveguide propagation loss coefficient, $\alpha$, simulated by FDTD.\cite{Tanaka03} $V_g$ and $\alpha$ are normalized by $c$ (speed of light) and $2\pi/a$, respectively.}
\end{figure}

To better understand the highly effective vertical confinement mechanism, let us now consider a hypothetical PhC slab cavity with $T = \infty$. The resulting structure may be viewed as a PhC fiber,\cite{Joannopoulos_book} thus one can define a waveguide dispersion in the $z$ direction. In Fig.~\ref{fig:fig3}(a), we show the waveguide dispersion of the dipole mode. It should be noted that these modes are not PBG-guided except for $k_z = 0$ point because a nonzero wavevector ($k_z > 0$) breaks the TE/TM symmetry and the original 2D PhC structure with $R = 0.35 a$ cannot have a complete PBG both for TE and TM.\cite{Joannopoulos_book} Thus, the guided modes with $k_z > 0$ are {\em inherently} lossy. Now we will show that the observed three resonant modes in Fig.~\ref{fig:fig1}(c) originate from these guided modes. In Fig.~\ref{fig:fig3}(a), we show intersection points between the dispersion curve and the three normalized frequencies ($\omega_n = a / \lambda$) of the resonant modes. We find that these points are almost equally arranged in the $k$ space, where $\triangle k_z$ indeed satisfies the Fabry-P\'{e}rot resonance condition, $\triangle k_z = \pi/T$; $\triangle k_z/(2\pi/a) = a/(2T) \approx 0.076$.\cite{JAP_seheon} Note that the group velocity ($V_g \equiv d\omega /dk$) of the fundamental dipole mode will approach zero as $T \rightarrow \infty$ and $k_z \rightarrow 0$ [See Fig~\ref{fig:fig3}(b)]. 

In view of this waveguide model, $Q_{\rm tot}$ of the fundamental dipole mode can be written as the sum of waveguide propagation loss and scattering loss at the two mirror facets such that\cite{JAP_seheon, Coldren_book} 

\begin{equation}
\frac{1}{Q_{\rm tot}} = \frac{V_g}{\omega} \left[ \alpha + \frac{1}{T} \log\left( \frac{1}{r_0^2} \right) \right].\label{eq:eq1}
\end{equation}
Here, $\omega$ is the angular frequency of the resonant mode and $\alpha$ is the waveguide propagation loss coefficient describing the imperfect horizontal photon confinement due to both the finite $x$-$y$ domain size and coupling into the higher-order slab modes.\cite{S_G_Johnson_99} As shown in Fig.~\ref{fig:fig3}(b), $\alpha$ varies as a function of $k_z$; it tends to diverge as $k_z \rightarrow 0$ due to the presence of a zero group velocity at $k_z = 0$.\cite{Kuramochi_05} $r_0$ is a reflection coefficient and $(1/T)\log(1/r_0^2)$ describes the scattering loss at the two mirror facets. Thus, $V_g \alpha / \omega$ and $V_g \log( 1/ r_0^2 ) / (T \omega)$ can be rewritten as $1/Q_{\rm horz}$ and $1/Q_{\rm vert}$, respectively.\cite{Qfactor} Now it is straightforward to show that $Q_{\rm vert}$ of the fundamental slab mode will grow indefinitely as $T \rightarrow \infty$ and $V_g \rightarrow 0$. The fact that the slow group velocity can enhance $Q$ of a resonant mode has been emphasized by Kim {\em et al.},\cite{JAP_seheon} who analyzed the ultra-high-$Q$ mode in a PhC linear cavity, and by Ibanescu {\em et al.},\cite{Ibanescu05} who used the anomalous zero group velocity point in an axially uniform waveguide to design a high-$Q/V$ cavity on a dielectric substrate. However, $Q_{\rm horz}$ will be bound by a finite value as $k_z \rightarrow 0$; $Q_{\rm horz}$ will approach the $Q$ of an ideal 2D dipole cavity (TE mode). Therefore, this simple analysis based on the waveguide dispersion can explain major features in $Q$ behavior observed in Fig.~\ref{fig:fig2}. 

\begin{figure}
\centering\includegraphics[width=8.4cm]{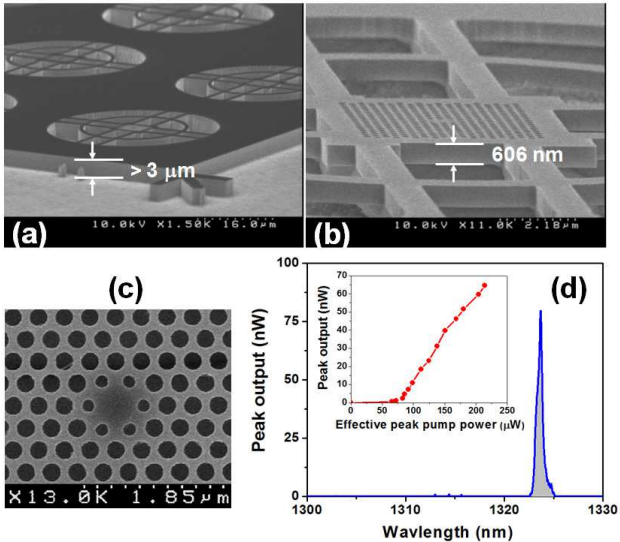}
\caption{\label{fig:fig4} (a-c) SEM images of PhC dipole lasers formed in a 606 nm InGaAsP slab. (a) Our dry-etching capability enables very deep ($>$3 $\mu{\rm m}$) and vertical etching. (b) A tilted image taken after selective wet-chemical etching. (c) A top view of the fabricated laser device. (d) Characteristics of the laser device.}
\end{figure}

In our experiment, PhC dipole mode cavities are fabricated in an InGaAsP slab with $T$ = 606 nm. Seven 60-\AA-thick  compressive-strained (1.0 \%) InGaAsP quantum wells emitting near 1.3 $\mu{\rm m}$ are embedded at the center of the slab, with 120-\AA-thick tensile-strained ($-0.3$ \%) 1.12 $\mu{\rm m}$ InGaAsP barriers in between. 240-nm-thick unstrained 1.12 $\mu{\rm m}$ InGaAsP is on top and bottom of the active layer and serve as a cladding. We use standard nano-fabrication processes including {\em e}-beam lithography (using hydrogen silsesquioxane as the resist), dry-etching to drill the PhC air-holes, and selective wet-chemical etching to undercut the InP sacrificial layer. To define deep and vertical air-holes, we use high-temperature ($190^{\circ}$C) Ar/${\rm Cl_2}$ chemically-assisted ion-beam etching (CAIBE). As shown in Fig.~\ref{fig:fig4}(a) and (b), our optimized CAIBE system produces very deep ($>$3 $\mu{\rm m}$) and vertical sidewalls, which are requisites to experimentally realize the theoretical $Q_{\rm tot}$ of 2,000 $\sim 3,000$. Fig.~\ref{fig:fig4}(b) and (c) show scanning electron microscope (SEM) images of fabricated laser devices.  

The fabricated lasers are photo-pumped at room-temperature with a 830 nm laser diode. The repetition rate of the pump laser is 1 MHz with a duty cycle of 2 \%. We use a 100$\times$ objective lens to focus the pump laser on to the cavity region. The same objective lens is used to collect the emitted laser light, which is fed into an optical spectrum analyzer. In Fig.~\ref{fig:fig4}(d), we present a light-in versus light-out ({\em L-L}) curve and a lasing spectrum for one example laser device. We confirm that the laser emission indeed comes from one of the degeneracy-split dipole modes\cite{Painter99} by comparing the emission wavelength (1323.7 nm) with that obtained by FDTD simulation. Assuming that about 20\% of actual incident pump power is absorbed in the slab, the effective threshold peak pump power is estimated to be 78 $\mu$W. 

Though the present work merely demonstrates an optically-pumped device, it is our hope that the thick-slab PhC cavity design will provide versatile routes toward a current-injection PhC laser. One feasible plan is to place the whole PhC slab cavity onto a metal substrate, where the metal may serve as both an electrical current pathway and a heat sink.\cite{S_H_Kim_11} An alternative is  to take advantage of the increased slab thickness, which enables more flexible design of the {\em p}-{\em i}-{\em n} doped layers and a current aperture structure. 

\bigskip

The authors would like to acknowledge support from the Defense Advanced Research Projects Agency under the Nanoscale Architecture for Coherent Hyperoptical Sources program under grant \#W911NF-07-1-0277 and from the National Science Foundation through NSF CIAN ERC under grant \#EEC-0812072. 



\begin{thebibliography}{10}
\bibitem{Qfactor}
$Q_{\rm tot}$ is defined by the decay rate of the total electromagnetic energy
  stored in the cavity such that $U(t) = U(0) \exp [ -\omega t /Q_{\rm tot} ]$.
  Then, the total radiation power ($\sim 1/Q_{\rm tot}$) can be decomposed into
  power radiated into the PhC slab ($\sim 1/Q_{\rm horz}$) and power radiated
  in the out-of plane direction ($\sim 1/Q_{\rm vert}$), therfore, $1/Q_{\rm
  tot} = 1/Q_{\rm horz} + 1/Q_{\rm vert}$.
\end{thebibliography}

\end{document}